\documentstyle[aps,floats,epsf,psfig,twocolumn]{revtex}
\tighten
\begin{document}
\draft 
\title{Neutron Scattering and the $B_{1g}$ Phonon in the Cuprates}
\author{$^{1}$T.P. Devereaux, $^{2}$A. Virosztek, and $^{2}$A. Zawadowski}
\address{$^{1}$Department of Physics, George Washington University, 
Washington, DC 20052}
\address{$^{2}$Institute of Physics and Research Group of the Hungarian 
Academy of Sciences, Technical University of Budapest, H-1521
Budapest, Hungary and Research Institute for Solid State Physics, 
P.O.Box 49, H-1525 Budapest, Hungary}
\date{\today}
\address{~
\parbox{14cm}{\rm 
\medskip
The momentum dependent lineshape of the out-of-phase
oxygen vibration as measured in recent neutron scattering measurements
is investigated.
Starting 
from a microscopic coupling of the phonon vibration to a local
crystal field, the
phonon lineshift and broadening is calculated as a function of transfered 
momentum in the superconducting state of YBa$_{2}$Cu$_{3}$O$_{7}$. For
the first time, the absolute magnitude of the shift is obtained as a 
function of momentum and is
found to be in excellent agreement with experiment.
It is shown that the anisotropy of the density of states,
superconducting energy gap, and the electron-phonon coupling are
all crucial in order to explain these experiments.}}
\maketitle

\narrowtext

Recently attention has focused on the observance of a resonance at an energy 
near 41 meV in the neutron scattering cross section obtained
in both optimally doped and underdoped
YBa$_{2}$Cu$_{3}$O$_{7-\delta}$\cite{keimer1,keimer2}. 
It has been speculated that this resonance feature
might hold the key to the pairing mechanism which underlies superconductivity
in the cuprate materials. As such, this feature has been lavished with
attention by many various theoretical proposals\cite{theories}.
Earlier measurements had suggested that this resonance was related to a
particular phonon vibration which occurs at nearly the same energy.
This phonon consists of the out of phase $c-$axis vibration of the
two oxygen O(2) and O(3) atoms in the CuO$_{2}$ plane.
However it was later clarified via kinematic analysis and
polarized neutron studies
that a magnetic and phononic contribution are both present and
could be separated.
Both peaks could be tracked as a function of temperature as well as scattered
neutron momentum. The phonon peak\cite{keimer2} 
has received a much smaller amount of attention in comparison
with the magnetic resonance.

Earlier Raman\cite{raman} and neutron scattering
\cite{neutron} experiments
have studied this particular phonon vibration in detail. 
Of all the Raman-active 
phonons measured in tetragonal superconductors, only this phonon
does not transform according to the full symmetry ($A_{1g}$) of the lattice.
Since this phonon obeys its own selection rules ($B_{1g}$) it can be
unambiguously identified. From its Fano profile, the magnitude of the
electron-phonon (e-ph) coupling seems to be particularly large for this phonon
compared to other phonons measured in Raman scattering. 
Indeed from the Raman point of view, this phonon has also been
lavished with attention\cite{marsiglio,normand,oka,dvz}. 
In Ref. \cite{marsiglio}, a phenomenological form for
the coupling was assumed and thus predictions concerning the absolute
magnitude of the lineshift could not be obtained. Moreover, 
a simplified momentum dependence of the coupling was assumed.
Only
Refs. \cite{normand,oka} and \cite{dvz} specified a microscopic source of the 
coupling.  In Refs. \cite{normand,oka} the coupling was a result of the 
buckling of the Cu-O plane. On the other hand, Ref. \cite{dvz}
has addressed this coupling as arising through a local crystal electric
field ${\bf E}$ which breaks the inversion symmetry locally around the 
Cu-O plane \cite{barisic}. Here the absolute value of the coupling
estimated from calculations of the electric field gave good agreement
to the results obtained from Raman scattering studies.
Neither theories made predictions about the momentum dependence of the
phonon lineshape.

Ref. \cite{keimer2} showed that
in the normal state the $B_{1g}$ phonon does not show any appreciable
dependence on the in-plane neutron's scattered momentum ${\bf q}$.
However in the superconducting state the phonon 
softened considerably for ${\bf q}=0$, in agreement with Raman
measurements, while the effect became rapidly less pronounced with increasing
$q$.  Along the $(1,1,0)$ direction in the Brillouin Zone (BZ),
the phonon hardened faster with increasing $q$ than along
the $(1,0,0)$ direction. While these findings were related to 
anisotropies of the energy gap, density of states, and the e-ph
coupling, an unambiguous determination of the importance of each
effect could not be made.
We therefore refocus attention on the phonon contribution to neutron
scattering in an effort to provide a theoretical framework to
better understand the contributions from various anisotropies.
Using a microscopic coupling theory developed in \cite{dvz}, 
the momentum dependence of the phonon lineshape will be used
to determine the interplay of the anisotropies of the energy gap, density of 
states, and the e-ph coupling. For the first time, an estimate for the
magnitude of the lineshape changes as a function of momentum
can be obtained.

Our starting point is 
a three-band model for the CuO$_{2}$ plane
with Cu-O hopping amplitude $t$ and O-O hopping amplitude $t^{\prime}$:
\begin{eqnarray}
&&H_{0}=\varepsilon\sum_{{\bf n},\sigma}b^{\dagger}_{{\bf n},\sigma}
b_{{\bf n},\sigma}
+t\sum_{{\bf n},\sigma,\pmb{$\delta$}}[P_{\pmb{$\delta$}}
b^{\dagger}_{{\bf n},\sigma}
a_{{\bf n},\pmb{$\delta$},\sigma}+h.c.]
\label{one}
\nonumber\\
&&+t^{\prime}
\sum_{{\bf n},\sigma}\sum_{\langle\pmb{$\delta$},\pmb{$\delta$}^\prime\rangle}
P^\prime_{\pmb{$\delta$},\pmb{$\delta$}^\prime}a^\dagger_{{\bf n},
\pmb{$\delta$},\sigma}a_{{\bf n},\pmb{$\delta$}^\prime,\sigma},
\end{eqnarray}
where $b^{\dagger}_{{\bf n},\sigma}$ creates an electron with spin
$\sigma$ at a copper lattice site ${\bf n}$, while 
$a_{{\bf n},\pmb{$\delta$},\sigma}$ 
annihilates an electron at one of the neighboring oxygen sites
${\bf n}+\pmb{$\delta$}/2$
determined by the unit vector $\pmb{$\delta$}$ assuming the four values,
$(\pm 1,0)$ and $(0,\pm 1)$. An oxygen atom between the two copper atoms at
sites ${\bf n}$ and ${\bf n}+\pmb{$\delta$}$ is labeled by either
$({\bf n},\pmb{$\delta$})$ or $({\bf n}+\pmb{$\delta$},-\pmb{$\delta$})$.
As in \cite{dvz}, $\varepsilon=E_d-E_p$ is the difference of the Cu and O site
energies and $P_{\pmb{$\delta$}}=\pm 1$
depending on whether the orbitals (with real wavefunctions) have the same
or opposite sign at the overlap region. Assuming Cu $d_{x^2-y^2}$ and
O $p$ orbitals $P_{-\pmb{$\delta$}}=-P_{\pmb{$\delta$}}$, and we can choose
$P_{(1,0)}=1$ and $P_{(0,1)}=-1$. Lastly, 
$P^{\prime}_{\pmb{$\delta$},\pmb{$\delta$}^\prime}$ denotes the
overlap sign between an
oxygen orbital at site ${\bf n}+\pmb{$\delta$}/2$ with a
neighboring oxygen orbital at site ${\bf n}+\pmb{$\delta$}^\prime/2$.
The O-O hopping is needed to give the right curvature and centering of
the observed Fermi surface. By our above
convention these overlaps take the values $P^{\prime}_{{\bf x},{\bf y}}=
P^{\prime}_{-{\bf x},-{\bf y}}=1,
P^{\prime}_{{\bf x},-{\bf y}}=P^{\prime}_{-{\bf x},{\bf y}}=-1$,
respectively. After Fourier transforming, the Hamiltonian now reads
$H^{0}=\sum_{{\bf k},\sigma} H^0_{{\bf k},\sigma}$, where
\begin{eqnarray}
H^0_{{\bf k},\sigma}=&&\varepsilon b^{\dagger}_{{\bf k},\sigma}b_{{\bf k},
\sigma} 
\label{two} \\
&&+\{ ib^{\dagger}_{{\bf k},\sigma}[a_{x,{\bf k},\sigma}t_x({\bf k})
-a_{y,{\bf k},\sigma}t_y({\bf k})]+h.c.\}\nonumber \\
&&+t^{\prime}({\bf k})[a_{x,{\bf k},\sigma}^{\dagger}a_{y,{\bf k},\sigma}+
h.c.],
\nonumber
\end{eqnarray}
with the prefactors
$t_{\alpha}({\bf k})=2t\sin(ak_{\alpha}/2),~~ 
t^{\prime}({\bf k})=-4t^{\prime}\sin(ak_{x}/2)\sin(ak_{y}/2).$
We can then diagonalize Eq. (\ref{two}):
$H^{0}_{{\bf k},\sigma}=\sum_{\beta}\epsilon_{\beta}({\bf k})
d_{\beta,{\bf k},\sigma}^{\dagger}
d_{\beta,{\bf k},\sigma},$
where $\beta$ is +,-, and 0 for the antibonding, bonding, and 
nonbinding bands, respectively. We consider only a reduced one band model
appropriate near half filling and take only the upper band into 
account\cite{upper}.
At this point the same procedure used in Ref. \cite{dvz} can be carried
through.
The electric fields at the O(2) and O(3) sites couple
linearly to their c-axis ion displacements leading to the following
e-ph coupling
\begin{eqnarray}
H_{el-ph}=eE{\bf \hat z}\cdot\sum_{{\bf n},\sigma}\{ {\bf u}_{x}(a{\bf n})
a_{{\bf n,x},\sigma}^{\dagger}a_{{\bf n,x},\sigma}+
\label{three}
\nonumber \\
{\bf u}_{y}(a{\bf n})
a_{{\bf n,y},\sigma}^{\dagger}a_{{\bf n,y},\sigma}\},
\end{eqnarray}
where $E$ is the $c-$axis aligned
electric field at the O(2) and
O(3) sites, neglecting orthorhombic distortions.
This can be rewritten in a momentum representation as
\begin{equation}
H_{el-ph}={1\over{\sqrt{N}}}\sum_{{\bf q,k},\sigma}g({\bf k,q})
d^{\dagger}_{{\bf k},\sigma}d_{{\bf k-q},\sigma}[c_{\bf q}+
c^{\dagger}_{-{\bf q}}],
\label{four}
\end{equation}
where $c_{\bf q}$ annihilates the $B_{1g}$ phonon mode with wave vector
${\bf q}$ and $g({\bf k,q})$ is the coupling constant of the mode to an
electron of wave vector ${\bf k}$. 

The coupling constant for the ${\bf q}=0$
$B_{1g}$ phonon was evaluated in \cite{dvz}. 
These results can be generalized for finite ${\bf q}$ as
\begin{eqnarray}
&&g({\bf k,q})=eE_{z}\sqrt{\hbar\over{2M_{O}N({\bf q})\omega_{B_{1g}}}} 
\times\label{five} \\
&&\biggl[\phi_{x}^{*}({\bf k})\phi_{x}({\bf k-q})e^{-iq_xa/2}(1+
e^{-iq_ya})-\nonumber \\
&&-\phi_{y}^{*}({\bf k})\phi_{y}({\bf k-q})e^{-iq_ya/2}(1+e^{-iq_xa})
\biggr].\nonumber
\end{eqnarray}
Here $M_{O}$ is the oxygen mass and $\omega_{B_{1g}}\sim 41$ meV is the
phonon frequency. The $\phi$ functions result from the diagonalization of the
three band model and are given as
\begin{equation}
\phi_{x,y}({\bf k})={\mp i\over{N_{\bf k}}}[t_{x,y}({\bf k})-t^{\prime}({\bf k})t_{y,x}({\bf k})/\epsilon({\bf k})]
\label{six}
\end{equation}
with the normalization factor
\begin{eqnarray}
&&N_{\bf k}={1\over{\epsilon({\bf k})}}
\bigl[(\epsilon({\bf k})^{2}-t^{\prime}({\bf k})^{2})^{2}+(\epsilon({\bf k})t_{x}({\bf k})-
\label{seven}
\nonumber \\
&&t_{y}({\bf k})t^{\prime}({\bf k}))^{2}+(\epsilon({\bf k})t_{y}({\bf k})-
t_{x}({\bf k})t^{\prime}({\bf k}))^{2}\bigr]^{1/2}.
\end{eqnarray}
Lastly the exponential factors and the other normalization $N({\bf q})$ 
comes from the phonon displacements. The eigenvalue and  
momentum dependent eigenvector for the $B_{1g}$ ionic vibration is solved
for the simple harmonic model of 3 atom unit cell of
Cu coupled via identical spring constants $K$
to both the O(2) and O(3) atoms under a uniform tension. Besides an 
acoustic mode and an in-phase O(2)-O(3) optical mode
($A_{1g}$),
the $B_{1g}$ mode is obtained with the dispersionless energy eigenvalue
$\omega^{2}_{B_{1g}}=2K/M_{O}$, and whose eigenvector is 
\begin{eqnarray}
&&\pmb{$\epsilon$}^{O(2,3)}_{B_{1g}}({\bf q})=
\pm{\bf \hat z}(1+e^{-iq_{y,x}a})/\sqrt{N({\bf q})},
\label{eight}
\\
&&N({\bf q})=4[\cos^{2}(q_{x}a/2)+\cos^{2}(q_{y}a/2)].
\nonumber
\end{eqnarray}
The
eigenvectors from this simplified model give an adequate agreement
to the eigenvectors
obtained via more complex lattice dynamical calculations\cite{private}. 

The e-ph coupling renormalizes the bare phonon propagator 
$D^{0}_{B_{1g}}
({\bf q},\omega)=2\omega_{B_{1g}}/[\omega^{2}-\omega_{B_{1g}}^{2}]$
such that the retarded propagator becomes
\begin{equation}
D_{B_{1g}}({\bf q},\omega+i\delta)={D^{0}_{B_{1g}}({\bf q},\omega)\over{1+D_{B_{1g}}^{0}({\bf q},\omega)
\Pi_{B_{1g}}({\bf q},\omega+i\delta)}},
\label{nine}
\end{equation}
where $\Pi_{B_{1g}}$ is the ${\bf q}$ and $\omega$ dependent polarization
\begin{eqnarray}
&&\Pi_{B_{1g}}({\bf q},\omega+i\delta)=\sum_{\bf k}\mid g({\bf k,q})\mid^{2}
\label{ten}\\
&&\times\biggl\{a_{+}({\bf k,q})[f(E_{\bf k})-f(E_{\bf k+q})]\nonumber\\
&&\times\biggl({1\over{\omega+i\delta+E_{\bf k}-E_{\bf k+q}}}-
{1\over{\omega+i\delta-E_{\bf k}+E_{\bf k+q}}}\biggr)\nonumber \\
&&+a_{-}({\bf k,q})[1-f(E_{\bf k})-f(E_{\bf k+q})]\nonumber \\
&&\times\biggl(
{1\over{\omega+i\delta-E_{\bf k}-E_{\bf k+q}}}-
{1\over{\omega+i\delta+E_{\bf k}+E_{\bf k+q}}}\biggr)\biggr\},
\nonumber
\end{eqnarray}
where $E_{\bf k}=\sqrt{\xi_{\bf k}^{2}+\Delta^{2}({\bf k})}$, 
$\xi_{\bf k}=\epsilon({\bf k})-\mu$ 
is the energy dispersion measured from the chemical potential $\mu$,
$f$ is the
Fermi function, and $a_{\pm}({\bf k,q})$ are coherence factors
\begin{equation}
a_{\pm}({\bf k,q})=1\pm {\xi_{\bf k}\xi_{\bf k+q}-\Delta({\bf k})
\Delta({\bf k+q})\over{E_{\bf k}E_{\bf k+q}}}.
\label{eleven}
\end{equation}

\begin{figure}
\hskip2.cm
\psfig{file=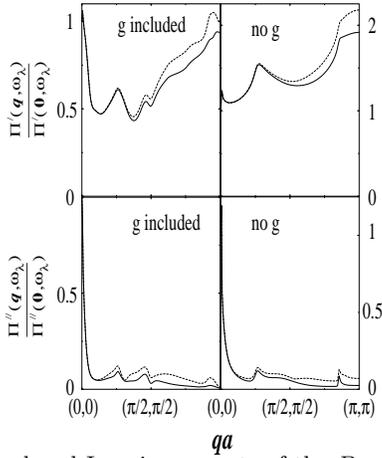,height=6.cm,width=5.cm,angle=0}
\caption[]{
Real and Imaginary parts of the $B_{1g}$ phonon self energy 
as a function of momentum transfer ${\bf q}$ along the BZ diagonal for
$T=T_{c}/2$. The solid lines in the figures correspond to the
$d_{x^{2}-y^{2}}$ energy gap and the long dashed lines correspond to the 
$\mid d_{x^{2}-y^{2}} \mid$ energy gap. The left panels represent the
evaluation of Eqs. (\ref{ten},\ref{eleven}) using the expression 
for the coupling constant $g$, Eq. (\ref{five}), while the right panels
correspond to using a momentum-independent coupling constant equal to $t$.}
\label{fig1}
\end{figure} 

This expression is identical to the one used in Ref. 
\cite{marsiglio}. 
However, all other uses of Eqs. (\ref{ten},\ref{eleven}) 
have used a momentum
independent coupling constant and have not specified a microscopic
mechanism for the electron-phonon interaction. 
We remark here this misses crucial information
of the interplay of symmetry of the phonon vibration and its coupling to
the electronic system. First of all, the use of ${\bf k}$-
independent coupling constant in Eqs. (\ref{ten},\ref{eleven}) 
corresponds to a particular case of a fully symmetric phonon ($A_{1g}$), 
transforming
as the identity representation. As a consequence
of the long wavelength Coulomb interaction, the ${\bf q}=0$ self energy
would be completely screened, leading to no net linewidth change for
${\bf q}=0$ phonons, as measured e.g. in Raman scattering 
measurements\cite{dvz,tpdphonons}. If a fully symmetric phonon has a 
momentum dependence, it still will be at least partially screened by
the long wavelength Coulomb interaction. However, there is no screening
for phonons of other symmetry. While
it appears that certain $A_{1g}$ phonons in the cuprates are weakly
temperature dependent and may indeed have their coupling to the
electron system screened out, the Ba $A_{1g}$ phonon as well as the 
O $B_{1g}$ vibration show
distinctive shifts as a function of temperature. This can only come about
if the phonons have a non-trivial momentum dependent coupling constant and 
therefore the use
of a momentum independent coupling constant is inappropriate for these
phonons\cite{raman}. 
Secondly, the neglect of the momentum dependence of the coupling
constant overestimates the size of the $B_{1g}$ phonon broadening
shift for ${\bf q}=0$. As with the electronic contribution to the Raman
response\cite{ers}, the coupling of the gap and the vertex which have the same
symmetry leads to a reduction of the phonon broadening in the superconducting
state $\sim (\omega/2\Delta)^{3}$ 
compared to the case when the gap and vertex
are of different symmetries $\sim \omega/\Delta$\cite{tpdphonons}. Lastly,
the momentum dependence is not separable into two functions of ${\bf k}$
and ${\bf q}$ respectively, as implied in Ref. \cite{marsiglio}.

\begin{figure}
\hskip2.cm
\psfig{file=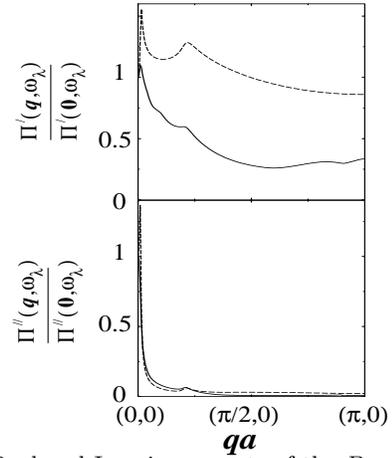,height=6.cm,width=5.cm,angle=0}
\caption[]{
Real and Imaginary parts of the $B_{1g}$ phonon self energy 
as a function of momentum transfer ${\bf q}$ along the BZ axis for
$T=T_{c}/2$. Both gap choices yield identical results.
The solid lines in the figures correspond 
to evaluating Eqs. (\ref{ten},\ref{eleven}) using the expression 
for the coupling constant $g$, Eq. (\ref{five}), while the dashed lines 
correspond to using a momentum-independent
coupling constant equal to $t$.}
\label{fig2}
\end{figure} 

Here we consider either an anisotropic $s-$wave or 
$d_{x^{2}-y^{2}}$ energy gap: $\Delta_{s}({\bf k})=\Delta_{0},
\mid[\cos(k_{x}a)-\cos(k_{y}a)]\mid/2,
~\Delta_{d}({\bf k})=\Delta_{0}[\cos(k_{x}a)-\cos(k_{y}a)]/2$.
While we believe that there is strong evidence that the energy gap
in the cuprates has $d_{x^{2}-y^{2}}$ symmetry, we consider the $s-$wave
case for illustrative purposes showing the effect of a sign change of
the energy gap. In the following we
choose $\varepsilon=1$eV, $t=1.6$eV, $t^{\prime}/t=0.45$ (which are
similar to the values chosen in Ref. \cite{oka}), $\Delta_{0}=30$meV 
and $2\Delta_{0}/k_{B}T_{c}=8$ for both energy gap choices, 
and $\omega = 41$meV. We adjust the chemical
potential to yield a filling $\langle n \rangle=0.875$ for both spins,
and set $T=T_{c}/2$.

In Fig. (1) we plot the real and imaginary parts of
$\Pi$ as a function of transfered momentum ${\bf q}$ along the (1,1,0)
direction for a $d_{x^{2}-y^{2}}$ and an anisotropic $s-$wave superconductor,
while Fig. (2) plots these same functions for ${\bf q}$
along the (1,0,0) direction. 
We have normalized both the real and imaginary parts of $\Pi$ to their
${\bf q}=0$ values. 
The three factors governing the ${\bf q}$ dependence of the $B_{1g}$
phonon's lineshape are: the joint density of states for phonon
scattering, the coherence factors, and the e-ph coupling.

We first consider features of the the polarization $\Pi$ which
are only due to kinematic constraints and band structure.
Starting from $q=0$, where Cooper pairs are broken equally
around the Fermi surface (FS), the pair-breaking becomes less
resonant as $q$ increases until two regions of the FS can
be connected. At the same time
the scattering processes at low temperatures which are most prominent are
those which can most closely satisfy the condition that
$\omega=\mid\Delta({\bf k})\mid+\mid\Delta({\bf k+q})\mid$ coming from
last term in Eq. (\ref{ten}). Both of these 
restrictions become less and less obeyed with increasing ${\bf q}$
and lead to a very rapid drop off of $\Pi$  away from
${\bf q}=0$ in either direction as seen in both figures regardless of the
energy gap symmetry. 
Moreover, as ${\bf q}$ approaches the zone boundary along the diagonal
(Fig. 1), both the real and imaginary parts of $\Pi$ rise
since regions of
the FS with large density of states (flat bands) are connected
by momentum transfer of $({\pi/a,\pi/a})$. 

\begin{figure}
\hskip2.cm
\psfig{file=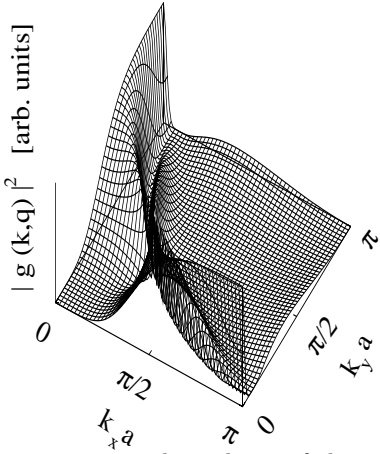,height=6.cm,width=5.cm,angle=270}
\caption[]{The momentum dependence of the coupling constant 
$\mid g({\bf k},{\bf q}=0)\mid^{2}$ as a function of momentum in
the BZ.}
\label{fig3}
\end{figure} 

Now if we consider the
anisotropy of the energy gap, we note that for
the anisotropic
$s-$wave case this rise is more pronounced than the $d-$wave case 
due to the change of sign of the gap in the coherence factor $a_{-}$ for
the $d-$wave gap. The rapid fall off of the
the data in Ref. \cite{keimer2} gives support for $d_{x^{2}-y^{2}}$
pairing. As we would expect, there is no
difference between $\Pi$ evaluated for the
two gaps for ${\bf q}$ transfer along the $(1,0,0)$ direction since
the coherence factor is the same for both energy gaps for scattering
near the FS.

We now consider the momentum dependence of the coupling constant.
In the Fig. 1 (left panel) and Fig. 2 
the solid line represents $\Pi$ calculated with the
coupling constant included while the right panel of Fig. 1
and the dashed lines in Fig. 2 are obtained by neglecting
the ${\bf k,q}$ dependence of the coupling constant.
We see that the results for both ${\bf q}$ directions are strongly
affected by the coupling constant and its inclusion cannot be neglected. 
In all cases the coupling constant
leads to a more rapid fall off of the polarization with increasing
${\bf q}$. This is due to the strong momentum dependence of the 
coupling constant $g({\bf k,q})$. For ${\bf q}=0$ the vertex varies
as $\sim[\cos(k_{x}a)-\cos(k_{y}a)](1-t^{\prime}({\bf k})/\epsilon({\bf k}))$
- it vanishes along the BZ diagonal and just off the FS when 
$t^{\prime}({\bf k})=\epsilon({\bf k})$. This is seen in Figs. 3, 4, and 5, 
which plot the vertex for ${\bf q}=(0,0), (\pi/a,0),$ and
$(\pi/a,\pi/a)$, respectively. Thus the vertex suppresses
scattering just off the FS plus scattering between BZ diagonals,
leading
to the rapid fall off of $\Pi$ with momentum transfer ${\bf q}$ as seen
in Figs. (1-2). This is in general agreement with the results from Ref.
\cite{keimer2}, supporting the importance of the coupling constant.

\begin{figure}
\hskip2.cm
\psfig{file=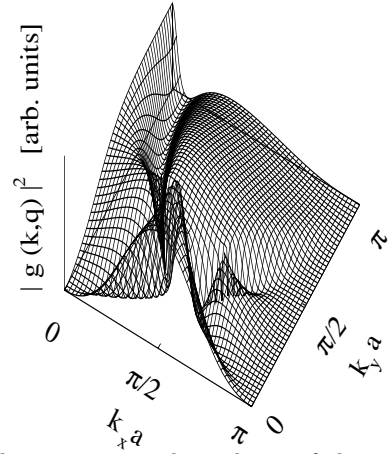,height=6.cm,width=5.cm,angle=270}
\caption[]{The momentum dependence of the coupling constant 
$\mid g({\bf k},{\bf q}=\pi/a,0)\mid^{2}$ as a function of momentum in
the BZ.}
\label{fig4}
\end{figure} 

\begin{figure}
\hskip2.cm
\psfig{file=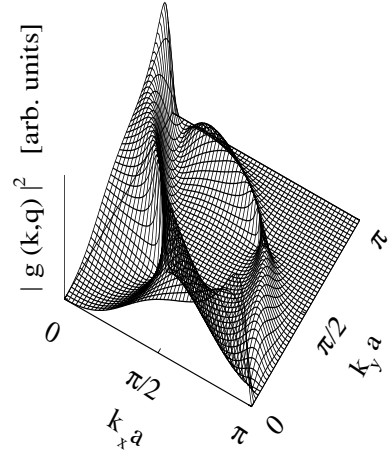,height=6.cm,width=5.cm,angle=270}
\caption[]{The momentum dependence of the coupling constant 
$\mid g({\bf k},{\bf q}=\pi/a,\pi/a)\mid^{2}$ as a function of momentum in
the BZ.}
\label{fig5}
\end{figure} 

Finally we consider the absolute magnitude for the hardening of the
$B_{1g}$ phonon as a function of ${\bf q}$. By specifying the microscopic
source of the e-ph coupling mechanism, we can for the first time provide
a numerical check of the strength of the coupling.
For this it is useful to
examine the neutron scattering cross section\cite{Squires} for 
scattering by the $B_{1g}$ phonon. 
The neutron scattering cross section can be written as
\begin{equation}
{d^{2}\sigma\over{d\Omega d\omega}} \sim
\mid F({\bf q})\mid^{2} S({\bf q},\omega),
\label{twelve}
\end{equation}
where $F$ is the form factor for scattering by the $B_{1g}$ phonon, 
$S={\rm Im}D$ (see Eq.(9))
is the phonon spectral function, and $\hbar \omega$ is the
difference in energy of the scattered and incident neutron.
As in Ref. 
\cite{keimer2}, we consider Umklapp scattering of the neutron by
the phonon for momentum transfers of ${\bf Q}+{\bf q}$, where ${\bf Q}=
(2\pi/a,0)$.  This determines the form factor as
\begin{equation}
\mid F({\bf q})\mid^{2}={[\cos(q_{x}a/2)+\cos(q_{y}a/2)]^{2}\over{
\cos^{2}(q_{x}a/2)+\cos^{2}(q_{y}a/2)}}.
\label{thirteen}
\end{equation}

Using $\omega_{B_{1g}}=348$cm$^{-1}$ 
and the magnitude of the Cu-O buckling, $eE_{z}=-0.8eV/
\AA$\cite{ladik}, we plot in Fig. 6 the results for the
cross section for a $d_{x^{2}-y^{2}}$ superconductor
for ${\bf q}$  along the diagonal and axis 
of the BZ, respectively. The e-ph coupling has been included,
and the parameters used are the same as for Fig. (2). In addition, a
constant phonon linewidth of 2.5 cm$^{-1}$ was included which represents
the intrinsic broadening of the phonon (i.e., due to anharmonic lattice
potentials) which is present even in the insulating state\cite{dvz}.
Once the spectral function was obtained the curve
was convoluted with a Gaussian of width
2 meV to mimic the experimental resolution of the apparatus used in
\cite{keimer2}. 

\begin{figure}
\hskip2.cm
\psfig{file=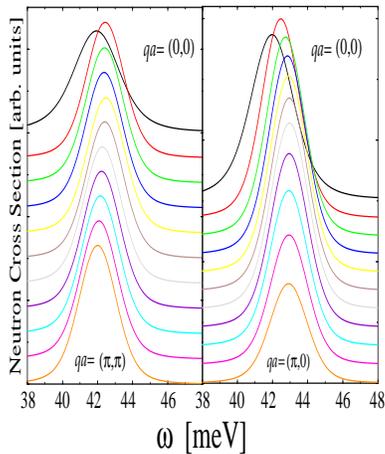,height=6.cm,width=5.cm,angle=0}
\caption[]{The neutron cross section as a
function of $\omega$ for different ${\bf q}$ values. The ${\bf q}$
values are chosen in equal increments along the diagonal (axis) of the
BZ for the left (right) panel, respectively.
Each curve has been offset for clarity.}
\label{fig6}
\end{figure} 

Our results for the frequency shift of the phonon for different momenta
are compared to the experimental data in Fig. (7) using the above
parameters. We note that the theory compares both qualitatively 
and quantitatively well with the data in two regards. First, the
theory predicts a hardening of the phonon
for increasing momentum transfers in both directions in agreement
with experiment (although data along the BZ diagonal is not very complete).
Further, a maximum hardening of roughly 0.9 meV observed for momentum transfer
along the BZ axis is in agreement with the
measured\cite{keimer2} hardening of $0.9\pm0.1$meV. The general behavior away
from the zone center (${\bf q}=0$) is reproduced as well.
As mentioned the
data along the BZ axis cannot be used to determined whether the energy gap
changes sign, but we note that if the anisotropic $s$ energy gap was used
it would predict a larger growth of the shift along the BZ diagonals than
displayed in Fig. (7).

\begin{figure}
\hskip2.cm
\psfig{file=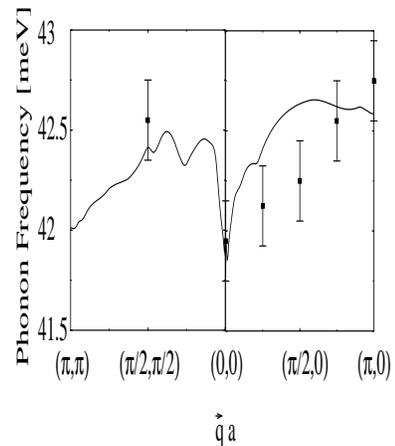,height=6.cm,width=5.cm,angle=0}
\caption[]{Comparison of the theory with the experimental data of
Ref. \cite{keimer2}.}
\label{fig7}
\end{figure} 

Three important details of the data are not captured by the theory however.
The theory gives a more abrupt fall off for the shift along the BZ axis 
than seen in the experiment, although a slightly larger error bars (as
seen in two points in Ref. \cite{keimer2}) would cover the discrepancy.
The experimental data show large changes for
the phonon frequency for $T=0.5T_{c}$ compared to $0.1T_{c}$ which the
theory cannot reproduce. 
Lastly, if the theory is used to calculate
the phonon frequency as a function of momentum in the normal state, the
phonon develops dispersion that roughly tracks the dispersion seen in the
superconducting state for momentum transfers larger than $\pi a/2$ in
either direction. This is also in disagreement with the experiment, which
shows a relatively
dispersionless phonon above $T_{c}$. 

These points may be indicative of the neglect of strong electronic
correlations. The theory does not include the effects of strong electronic
correlations as the electronic parameters are taken in accord with local
density approximation values\cite{oka}. This may crucially affect the
momentum dependence of the lineshape as the momentum vector maps out
the band dispersion as well as the shape of the Fermi surface. The strong
local Coulomb repulsion would lead to a smaller renormalized bandwidth
and thus a less rapid dependence of $\Pi$ with momentum. The theory could
be refined once parameter choices for the band structure are determined
via fitting to experimentally determined band dispersion and Fermi surfaces.
This remains to be investigated. 
The strong temperature dependence of the
phonon even below $T_{c}$ remains puzzling.
All theory curves are essential $T-$ independent
below $T=0.5T_{c}$ as the energy gap is completely established. 
The phonon lineshape changes below $T_{c}$
are due to the opening of a phonon decay channel by breaking Cooper pairs
near ${\bf q}=0$ and are minimized by momentum and energy conservation
at larger momentum transfers. Additional scattering via, e.g.
impurities or spin fluctuations, open channels for phonon
scattering at small ${\bf q}$ in the normal state as well which may counteract
the dispersion given by $\Pi$.  However, this cannot yield a
strong temperature dependent scattering substantially below $T_{c}$.

In summary we have investigated contributions to the momentum dependence
of the $B_{1g}$ phonon resulting
from anisotropies of the density of states, energy gap, and e-ph
coupling constant. Using a microscopic theory for the origin of the
e-ph coupling as given in \cite{dvz}, for the first time 
the theory's prediction for the 
dependence of the lineshape of the phonon as a function of
${\bf q}$ for various directions in the BZ is in qualitative agreement with
the results of Ref. \cite{keimer2}, while the
magnitude of the hardening of the phonon is in quantitative agreement. 
It is crucial that the anisotropies of all quantities are taken into account.

\section*{Acknowledgments}
T.P.D. would like to acknowledge helpful conversations with N. Bulut,
D. Scalapino, R. T. Scalettar, T. Fong, D. Reznik and B. Keimer. 
Acknowledgment is made
to the Donors of The Petroleum Research Fund, administered by the 
American Chemical Society, for partial support of this research.
This work was supported by the Hungarian National Research Fund under
Grant Nos. OTKA T020030, T016740, T021228/1996, T024005/1997,
and by the US-Hungarian Joint Fund No. 587.


\begin{references} 

\bibitem{keimer1}
H. A. Mook {\it et al.}, Phys. Rev. Lett. {\bf 70}, 3490 (1993);
H. F. Fong {\it et al.}, Phys. Rev. Lett. {\bf 75}, 316 (1995);
Phys. Rev. B {\bf 54}, 6708 (1996); Phys. Rev. Lett.
{\bf 78}, 713 (1997).

\bibitem{keimer2}
D. Reznik {\it et al}, Phys. Rev. Lett. {\bf 75},
2396 (1995).

\bibitem{theories}
N. Bulut and D. J. Scalapino, Phys. Rev. B {\bf 47},
3419 (1993); E. Demler and S.-C. Zhang, Phys. Rev. Lett. {\bf 75},
4126 (1995); D. Z. Liu {\it et al.}, Phys. Rev. Lett. {\bf 75},
4130 (1995); I. I. Mazin and V. M. Yakovenko, Phys. Rev. Lett. {\bf 75}, 
4134 (1995); L. Yan {\it et al.}, Phys. Rev. Lett. {\bf 78}, 3559 (1997).

\bibitem{raman}
B. Friedl {\it et al.}, Phys. Rev. Lett. {\bf 65}, 915 (1990);
E. Altendorf {\it et al.}, Phys. Rev. B {\bf 47}, 8140 (1993).

\bibitem{neutron}
H. A. Mook {\it et al.}, Phys. Rev. Lett. {\bf 65}, 2712 (1990);
K. Pyka {\it et al.}, Phys. Rev. Lett. {\bf 70}, 1457 (1993).

\bibitem{marsiglio}
R. Zeyher and G. Zwicknagl, Z. Phys. B {\bf 78}, 175 (1990);
F. Marsiglio, Phys. Rev. B {\bf 47}, 5419 (1993); M. E. Flatt\'e, Phys. Rev.
Lett. {\bf 70}, 658 (1993); C. Jiang and C. Carbotte, Phys. Rev B
{\bf 50}, 9449 (1994); A. Bill, V. Hizhnyakov, and E. Sigmund, Phys. Rev.
B {\bf 52}, 7637 (1995); Journ. of Supercon. {\bf 9}, 493 (1996).

\bibitem{normand}
B. Normand {\it et al.}, J. Phys. Soc. Japan {\bf 64}, 3903
(1995); S. Y. Savrasov and O. K. Andersen, Phys. Rev. Lett. {\bf 77}, 4430
(1996).

\bibitem{oka}
O. K. Andersen {\it et al.}, Journ. of Low Temp. Phys. {\bf 105}, 285 (1996).

\bibitem{dvz} T. P. Devereaux {\it et al.}, Phys. Rev.
B {\bf 51}, 505 (1995); Sol. State Commun. {\bf 108}, 407 (1998).

\bibitem{barisic}
S. Bari\v si\'c and I. Batisti\'c, Europhys. Lett. {\bf 8}, 765 (1989).

\bibitem{upper}
The energy of the upper band follows as
$$
\epsilon({\bf k})=s_{+}({\bf k})+s_{-}({\bf k})+\varepsilon/3,$$
$$s_{\pm}({\bf k})=(r({\bf k})\pm\sqrt{q^{3}({\bf k})+r^{2}({\bf k})})^{1\over{3}}$$
$$q({\bf k})=-{1\over{3}}[t_{x}^{2}({\bf k})+t_{y}^{2}({\bf k})
+t^{\prime 2}({\bf k})]
-\varepsilon^{2}/9 $$
$$r({\bf k})={\varepsilon\over{6}}[t_{x}^{2}({\bf k})+
t_{y}^{2}({\bf k})-2t^{\prime 2}({\bf k})]$$
$$-t^{\prime}({\bf k})t_{x}({\bf k})t_{y}({\bf k})
+\varepsilon^{3}/27.$$
Since $q^{3}({\bf k})+r^{2}({\bf k})$ is negative for all 
${\bf k}$, $s_{+}^{*}({\bf k})=s_{-}({\bf k})$ 
and the energy is of course real.

\bibitem{private}
T. Fong, private communication.

\bibitem{tpdphonons}
T. P. Devereaux, Phys. Rev. B {\bf 50}, 10287 (1994).

\bibitem{ers}
T. P. Devereaux {\it et al.}, Phys. Rev. Lett. {\bf 72}, 396 (1994);
Phys. Rev. B {\bf 51}, 16336 (1995).

\bibitem{Squires} see eg. G. L. Squires, {\it Introduction to the Theory
of Thermal Neutron Scattering} (Dover, New York, 1996).

\bibitem{ladik}
$\omega_{B_{1g}}$ is slightly smaller than the value used to 
fit the $B_{1g}$ Fano profile in Ref. \cite{dvz}. 
The value for the electric field used is
in rough agreement with the value obtained from ab initio 
Hartree-Fock cluster calculations in
J. Li and J. Ladik, Sol. State Commun. {\bf 95}, 35 (1995).
 
\end{references}
\end{document}